# Notes on the second quantization in quantum electrodynamics

V. A. Golovko

Moscow Polytechnic University

Bolshaya Semenovskaya 38, Moscow 107023, Russia

E-mail: fizika.mgvmi@mail.ru

Abstract

It is demonstrated that the second quantization which is the basis of quantum electrodynamics is introduced without sufficient grounds and even logically inconsistently although it yields extremely accurate predictions that are in excellent agreement with experiment. The physical essence hidden behind the second quantization is discussed as well.

In Ref. [1] a new formulation of quantum electrodynamics (QED) is proposed in which the second quantization is absent and the electronic and electromagnetic fields are ordinary *c*-numbers. It is pointed out in [1] that the second quantization is introduced in QED without sufficient grounds and even logically inconsistently. In the present notes this question is discussed in more detail.

The second quantization was invented by Dirac. When considering assemblies of particles in his textbook [2] Dirac always takes as a ket for the assemblies the product of kets for each particle by itself $|a_1\rangle|b_2\rangle|c_3\rangle...|g_n\rangle$ without mentioning anywhere that such a representation holds only for noninteracting particles. All real particles interact with one another but Dirac does not at all discuss how the interaction might be taken into account. Dirac's ideas are considered in more detail by Schweber [3]. He starts from Eq. (26) of Sec. 6c

$$\Psi_{n_1 n_2 ... n_N}(q_1, q_2, ..., q_N) = \psi_{n_1}(q_1)\psi_{n_2}(q_2)...\psi_{n_N}(q_N) \qquad (1)$$

in a slightly modified notation. Here $q_i$ denotes the set of coordinates and the spin variable of the *i*th particle if the particles have spin, $n_i$ is the set of its quantum numbers, and $N$ is the number of particles in the system. Schweber just as Dirac does not mention that the wave function of the type (1) can describe only noninteracting particles (the function must be symmetrized or antisymmetrized). It would seem that the interaction between the particles can be taken into account in this method because in Eq. (110) of Sec. 6h Schweber writes

$$H = -\frac{\hbar^2}{2m}\int d\mathbf{x}\,\psi^*(\mathbf{x})\nabla^2\psi(\mathbf{x}) + \tfrac{1}{2}\int d\mathbf{x}\int d\mathbf{x}'\,\psi^*(\mathbf{x}')\psi^*(\mathbf{x})V(\mathbf{x},\mathbf{x}')\psi(\mathbf{x})\psi(\mathbf{x}'), \qquad (2)$$

where the last term must describe the interaction. In actual truth, this Hamiltonian is devoid of sense because the action of the operator $\psi(\mathbf{x})$ is defined only if the wave function is of the form (1) whereas the wave function cannot be of this form if there is an interaction between the particles.

The fact that one is dealing with systems of noninteracting particles in this method is clearly stated in §§ 64 and 65 of Ref. [4] devoted to the second quantization. It is instructive to remark that in previous editions of this textbook written when Landau was alive no mention was made of the situation where the particles would interact with one another, just as in [2] and [3]. After Landau's death his co-workers, trying nevertheless to say something about interacting particles, write in the revised edition [4], § 64, that in a system of interacting particles the occupation numbers $N_1$, $N_2$, … are no longer conserved and one can consider only the probability distribution of various values of the occupation numbers. The last statement made without any proof and even without any explanation raises many questions. Firstly, how can one define the single-particle states necessary for the second quantization if such states are absent in the interacting system? Secondly, it is necessary to prove that the single-particle states defined somehow or other satisfy all requirements and relations valid for the noninteracting systems. Thirdly, how can one find the above probability distribution of the various values of the occupation numbers? Fourthly, to conserve the essence of the method one should employ averaged occupation numbers $\overline{N}_1$, $\overline{N}_2$, …instead of $N_1$, $N_2$, … . In this connection the question arises as to whether the averaged occupation numbers are strictly equal to integers and why. If they are not integers, the method fails. As an example, we consider a Fermi system, in which case the numbers $\overline{N}_i$ should take only the values 0 or 1. If $\overline{N}_i = 1$, we should have $N_i > 1$ and $N_i < 1$ in the above probability distribution. How can the values $N_i > 1$ be reconciled with the Pauli exclusion principle? To obtain $\overline{N}_i = 0$ we should have $N_i > 0$ and $N_i < 0$. What do the negative occupation numbers signify? As a result we see that any attempt to extend Dirac's second quantization to the systems of interacting particles leads only to difficulties and contradictions.

At the same time there exists a strict mathematical method valid for the systems of interacting particles as well [5] (see also [6]). Let there be a complete set of orthonormal single-particle functions $\psi_n(q)$ with the same notation as in (1). Any function can be expanded into an infinite series in terms of $\psi_n(q)$ so that

$$\Psi(q_1, q_2, ..., q_N, t) = \sum_{n_1} \sum_{n_2} ... \sum_{n_N} c(n_1, n_2, ..., n_N, t) \psi_{n_1}(q_1) \psi_{n_2}(q_2) \cdots \psi_{n_N}(q_N). \qquad (3)$$

It should be emphasized that here we have an infinite series instead of (1).

We assume the Schrödinger equation of the form

$$i\hbar \frac{\partial \Psi}{\partial t} = \left[ \sum_{k=1}^{N} H_0(q_k) + \sum_{k \neq j}^{N} W(q_k, q_j) \right] \Psi, \qquad (4)$$

where $H_0(q_k) = -(\hbar^2/2m)\nabla_k^2 + U(q_k)$, $U(q_k)$ is the potential energy of the $k$th particle in an external field, and $W(q_k, q_j)$ the energy of interaction between the $k$th and $j$th particles. Substituting (3) into (4), multiplying on the left by $\psi^*_{m_1}(q_1)\psi^*_{m_2}(q_2)...\psi^*_{m_N}(q_N)$ and integrating with respect to $q_1, q_2, …, q_N$ we obtain a set of equations for $c(n_1, n_2, …, n_N, t)$. Instead of $c(n_1, n_2, …, n_N, t)$ it is convenient to introduce $c(N_1, N_2, …, t)$ where $N_1, N_2, …$ are the occupation numbers of the states $\psi_n(q)$ so that $|c(N_1, N_2, …, t)|^2$ is the probability of finding $N_1$ particles in the first state, $N_2$ particles in the second, and so on. The resulting equation can be written as ([5], Sec. 118)

$$i\hbar \frac{d}{dt} c(N_1, ..., N_m, ..., N_{m'}, ..., N_n, ..., N_{n'}, t)$$

$$= \sum_{n,m} \sqrt{N_m(N_n+1)} H_{mn} \, c(N_1,...,N_m-1,...,N_{m'},...,N_n+1,...,N_{n'},t)$$

$$+ \tfrac{1}{2} \sum \sqrt{N_m N_{m'}(N_n+1)(N_{n'}+1)} \, W_{mm',nn'} \, c(N_1,...,N_m-1,...,N_{m'}-1,...,N_n+1,...,N_{n'}+1,t) , \quad (5)$$

in which

$$H_{ln} = -\frac{\hbar^2}{2m}\int \psi_l^*(q)\nabla^2\psi_n(q)dq + \int \psi_l^*(q)U(q)\psi_n(q)dq , \tag{6}$$

$$W_{mm',nn'} = \int \psi_m^*(q)\psi_{m'}^*(q')W(q,q')\psi_n(q)\psi_{n'}(q')dqdq' . \tag{7}$$

Introducing creation $\mathbf{a}_n^*$ and annihilation $\mathbf{a}_n$ operators according to Eqs. (118.12) or (118.24) of [5] Eq. (5) can shortly be written as

$$i\hbar \frac{dc(N_1,N_1,...,t)}{dt} = \mathbf{H}c(N_1,N_1,...,t) , \tag{8}$$

where

$$\mathbf{H} = \sum_{m,n} \mathbf{a}_m^* H_{mn} \mathbf{a}_n + \tfrac{1}{2} \sum_{m,m'} \sum_{n,n'} \mathbf{a}_m^* \mathbf{a}_{m'}^* W_{mm',nn'} \mathbf{a}_n \mathbf{a}_{n'} . \tag{9}$$

To elucidate the meaning of Eq. (5) let us consider an example. When treating an ordinary differential equation one sometimes looks for a solution to the equation in terms of a series $f(x) = \Sigma \, a_n x^n$. Upon placing the series in the equation one arrives at a recurrence relation for the unknown coefficients $a_n$. Usually, such recurrence relations are of little use enabling one to calculate only first several coefficients $a_n$. In actual fact, Eq. (5) is a recurrence relation for the coefficients $c(N_1, N_2, ..., t)$. In the general case it is impossible to obtain concrete results from Eq. (5), the more so as one must sum the infinite series of (3). Equation (5) can be treated only approximately. This consideration demonstrates that Dirac's second quantization is a simplified and approximate version of the strict method in which one retains only several terms connected by symmetry or antisymmetry in the infinite series of (3).

A word should be said about Ref. [6]. In that reference Heisenberg considers the second quantization strictly without any approximation. He assumes that the wave function regarded as a $q$-number satisfies the commutation relations

$$\psi(P)\psi^*(P') - \psi^*(P')\psi(P) = \delta(P-P'),$$

$$\psi(P)\psi(P') - \psi(P')\psi(P) = 0, \quad \psi^*(P)\psi^*(P') - \psi^*(P')\psi^*(P) = 0 . \tag{10}$$

Next, he develops the wave function in a suitably chosen set of orthogonal functions $u_r(P)$:

$$\psi = \sum_r a_r u_r(P), \quad \psi^* = \sum_r a_r^* u_r(P) . \tag{11}$$

It follows from (10) that the coefficients $a_r, a_r^*$ satisfy the commutation relations

$$a_r a_s^* - a_s^* a_r = \delta_{rs}, \quad a_r a_s - a_s a_r = 0, \quad a_r^* a_s^* - a_s^* a_r^* = 0 \qquad (12)$$

(see § 10 of Appendix of [6]). All these relations are fully analogous to the ones written down, for example, in § 64 of [4] where only noninteracting particles are implied (see above). Heisenberg demonstrates, however, in § 11 that this strict formulation of the second quantization leads to the same recurrence relation as in (5) which is of little use as mentioned above [in the present context Eq. (8) is identical with (5)]. It is interesting to note that this Heisenberg's book is practically never cited when considering the second quantization.

There exists another approach in introducing the second quantization [7,8]. The approach is treated in detail by Bogoliubov and Shirkov [8]. One starts from quantum mechanical consideration of a harmonic oscillator where one can introduce creation and annihilation operators, which is well-known from textbooks on quantum mechanics. Thereupon one <u>postulates</u> by analogy that field functions can be regarded as operators and can be expressed linearly in terms of the creation and annihilation operators. The commutation rules for them are established according to a "correspondence principle". A separate postulate is required for fields describing particles of half-integral spin which are to be subject to Fermi-Dirac quantization. Here again we see that one leans heavily upon noninteracting particles, upon noninteracting harmonic oscillators in this approach, and thereupon puts forward some hypotheses without any justification. It should also be added that in all above references the second quantization is introduced in the framework of non-relativistic quantum mechanics, implying the non-relativistic Schrödinger equation for many bodies in particular, and the results obtained are tacitly utilized in relativistic theories without any explanation over again. In this connection it will be recalled that there does not exist a strictly formulated and generally accepted Dirac equation for many bodies, an equation analogous to the Schrödinger equation for many bodies, and relativistic many-body problems are treated with use made of one or other of approximations. By the way, the relativistic Dirac equation with a harmonic oscillator potential has no eigenvalue at all, which is completely different from the Schrödinger equation [9, p. 131], and it looks illogical to use just the last equation in order to sanctify the second quantization in the relativistic case too.

It was established by the trial-and-error method during two decades when QED was constructed that the second-quantization apparatus developed by Dirac without strict substantiation excellently works and gives extremely accurate predictions that are in exceptional agreement with experiment provided the apparatus is supplemented with mathematical methods frequently found empirically without strict proofs. This indicates that there is something hidden behind Dirac's second quantization, something physical that should be extracted from it. The essence of the method of the second quantization as a convenient mathematical tool in QED is discussed in Ref. [1]. It follows from [1] that the physical kernel of the method corresponds to the fact that the electron must be described by two *c*-number bispinors satisfying two mutually connected Dirac equations. The second bispinor is relevant to a charge of opposite sign which is required in order to fully compensate repulsive electric forces in the electronic cloud.

The success of the second quantization in QED can be explained as follows. QED leans essentially upon perturbation theory. This theory in the zeroth approximation implies noninteracting particles. Consequently, one can start with considering wave functions describing the noninteracting particles as in (1), which is tacitly done by all authors. Therefore the use of the second quantization is closely related to perturbative methods. In this connection the question arises as to the applicability of the second quantization in physics of strong interactions where the perturbative methods do not work. It may be added that doubt on the applicability of QED methods and thereby of the second quantization in the realm of strong interactions was cast by Bogoliubov and Shirkov [8, Sec. 14.1].